\documentclass{ws-rv9x6}
\usepackage{subfigure}     
\usepackage{ws-rv-van}     
\makeindex

\makeatletter
\newcommand{\raisemath}[1]{\mathpalette{\raisem@th{#1}}}
\newcommand{\raisem@th}[3]{\raisebox{#1}{$#2#3$}}
\makeatother

\begin{document}

\chapter[Exact record and order statistics of random walks via first-passage ideas]{Exact record and order statistics of random walks via first-passage ideas}\label{ra_ch1}

\author[G. Schehr and S. N. Majumdar]{Gr\'egory Schehr and Satya N. Majumdar}

\address{Laboratoire de Physique Th\'eorique et Mod\`eles Statistiques (LPTMS), Univ. Paris-Sud, CNRS, 91405 Orsay Cedex, France}

\begin{abstract}
While records and order statistics of independent and identically distributed (i.i.d.) random variables $X_1, \cdots, X_N$ are fully understood, much less is
known for strongly correlated random variables, which is often the situation encountered in statistical physics. Recently, it was shown, in a series of works, that one-dimensional random walk (RW) is an interesting laboratory where the influence of strong correlations on records and order statistics can be studied in detail. We review here recent exact results which have been obtained for these questions about RW, using techniques borrowed from the study of first-passage problems. We also present a brief review of the well known (and not so well known) results for records and order statistics of i.i.d. variables.  

\end{abstract}
\body

\section{Introduction}

Records and order statistics are by now a longstanding issue in the fields of engineering \cite{Gumbel}, finance \cite{EKM97} or environmental
sciences \cite{KPN02} where extreme events might have drastic consequences. Indeed, in these contexts, the statistics of extremes have practical applications which include the prediction of probability distributions of extreme floods, the amounts of large
insurance losses, equity risk, the size of freak waves, mutation events during evolution, extreme
statistics of time series, etc. These notions are very popular in our societies as, for instance, one always hears and reads, in the media, about record breaking events. This is especially true for sports, where world records are always special and noteworthy \cite{Gembris}.

More recently, it was realized that records and order statistics play a crucial role in statistical physics. Hence, there has
been a surge of interest for these questions in the physics literature. If one considers a discrete time series $X_1, \cdots, X_N$, where $X_i$'s might represent daily temperatures in a given city or the stock prices of a company, a record happens at time $k$ if the $k$-th entry is larger than all previous entries $X_0, \cdots, X_{k-1}$ (see Fig. \ref{fig_record} left). One is naturally led to ask the following questions: (a) how many records occur in time $N$ ? (b) how long does a record survive ? what is the longest or shortest age of a record ? Such questions and related ones have found applications in various physical situations ranging     
from domain wall dynamics \cite{ABBM}, spin-glasses \cite{Sibani} and random walks \cite{MZ2008,satya_leuven,sanjib,WMS2012,MSW2012} to avalanches \cite{LDW09}, models of stock prices \cite{WBK2011,WMS2012} or the study of global warming \cite{RP2006,WK2010} and also in evolutionary biology \cite{krugjain,franke} (see Ref. \cite{gregor_review} for a recent review). 

Another interesting question about this sequence concerns the fluctuations of the ordered sequence (so called order statistics) obtained by arranging the values of $X_i$ by decreasing order of magnitude, $M_{1,N} > M_{2,N} > \cdots > M_{N,N}$, $M_{k,N}$ being the $k$-th maximum of this sequence. Questions related to the statistics of the first maximum, $X_{\max}=M_{1,N}$ have emerged in various areas of physics ranging from disordered systems \cite{JPBM,PLDCecile,Dahmen} and fluctuating interfaces \cite{Shapir,GHPZ,Satya_Airy1,Satya_Airy2,SOS_Airy} to stochastic processes\cite{sire}, random matrices \cite{TW} and many others. While the statistics of the extremum $X_{\rm max}$ is important another natural question is: is this extremal 
value isolated, i.e., far away from the others, or is there many other 
events close to them? Such questions have led to the study of the density of 
states of near-extreme events \cite{SM07,SMR2008}. Order statistics is a natural way to  
characterize this phenomenon of crowding of near-extreme 
events. A set of useful observables that are naturally 
sensitive to the crowding of extremum are the gaps between 
the consecutive ordered maxima: $d_{k,N} = M_{k,N} - M_{k+1,N}$ denoting
the $k$-th gap. Such questions came up in several physical contexts, in particular in the study
of the branching Brownian motion \cite{BD2009,BD2011} and also for $1/f^\alpha$ signal \cite{MOR2011}, and more recently for random walks \cite{SM12,MMS13}. 

Records and order statistics of i.i.d. random variables are now perfectly well understood \cite{Gumbel,ABN1992,DN2003}, and we shall briefly review below the main results in this case. The record statistics when the entries $X_i$'s have a non-identical distributions but still retain their independence were also  studied in Ref. \cite{ballerini,Krug_records,Eliazar_records}, in the so called Linear Drift Model. On the other hand the order statistics of weakly correlated random variables reduce, to a large extent, to the case of i.i.d. random variables. However, much less is known for the difficult case where $X_i$'s are strongly correlated, which turns out to be the case of interest in many problems of statistical physics. Recently, it was shown that one-dimensional random walk (RW) is a non-trivial instance of a set of strongly correlated variables for which exact results for records \cite{MZ2008,WMS2012,MSW2012} and order statistics \cite{SM12,MMS13} can be obtained. In this paper, we review the main body of these results, which have been obtained, to a large extent, by methods and ideas stemming from first passage problems (for a review on this topic see \cite{Redner_book,Satya_review,Bray_review}).  

The paper is organized as follows. In section 2, we first focus on records statistics while we focus on order statistics in section 3. In each section, we first present a brief overview of well known, and not so well know, results for i.i.d. random variables. This is then followed by the review of results recently obtained for RW.

\section{Record statistics}

\subsection{Record statistics of i.i.d. random variables}

\begin{figure}
\centering
\includegraphics[width = 0.6\linewidth]{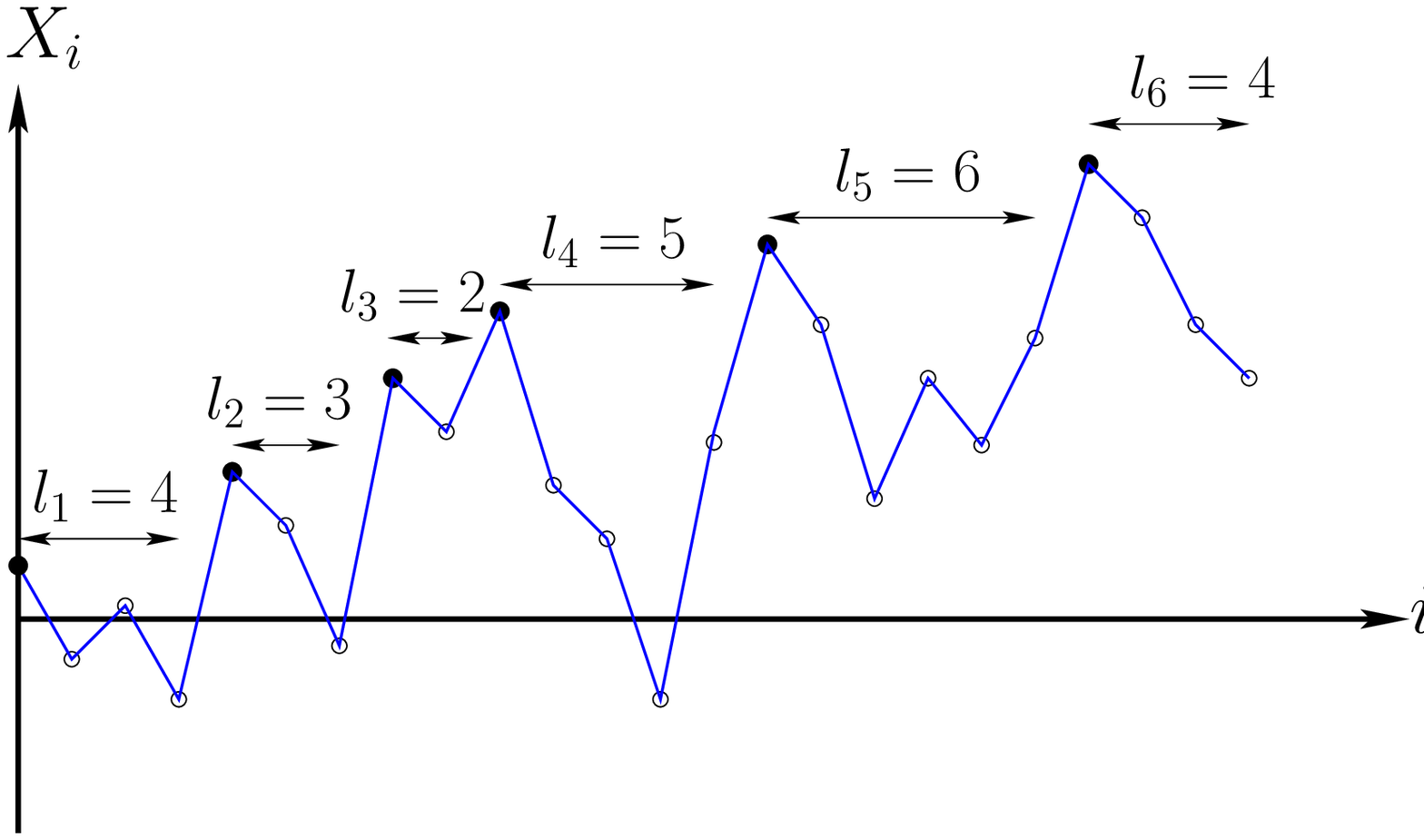}\includegraphics[width = 0.4\linewidth]{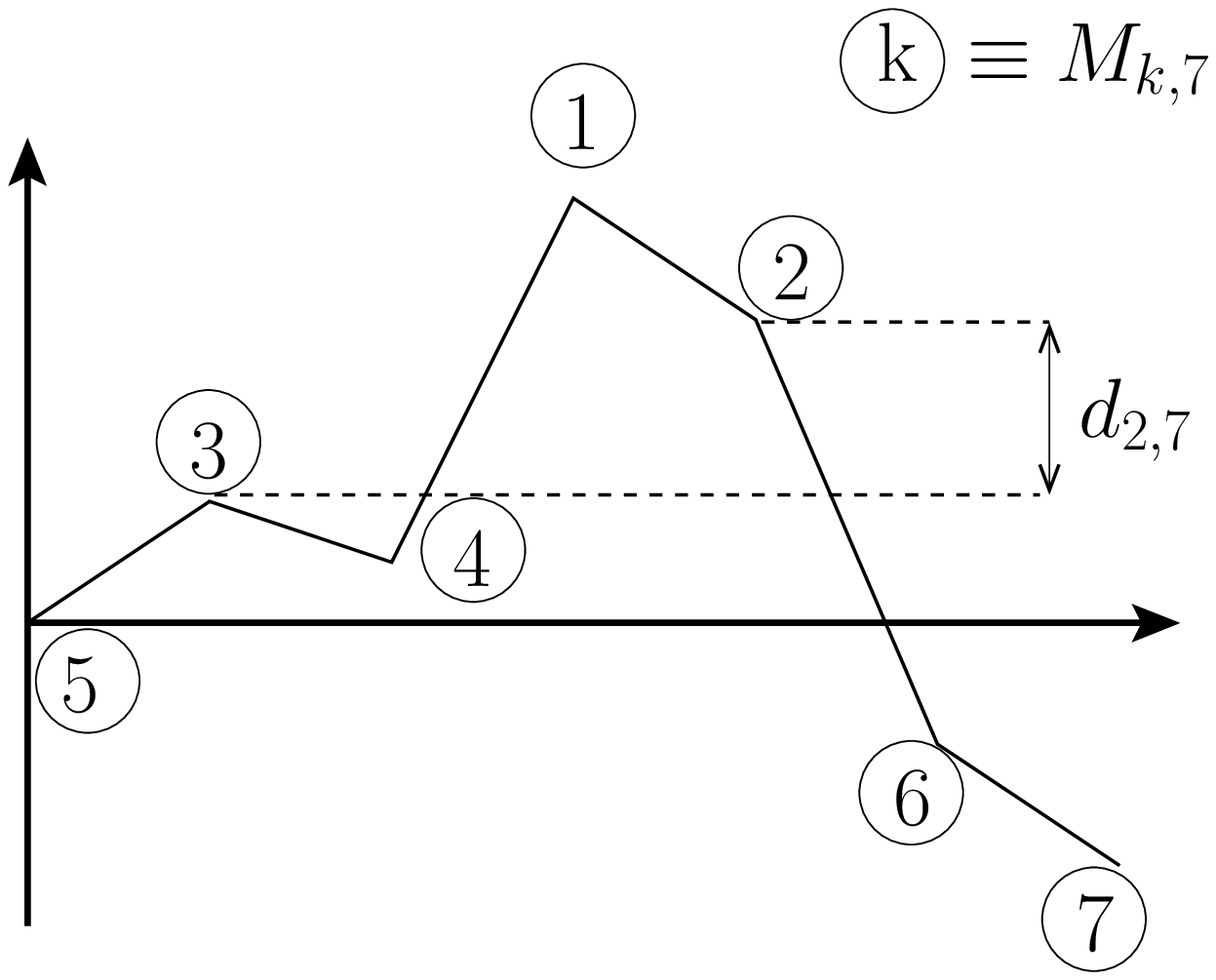}
\caption{{\bf Left}: One realization of $N=24$ random variables $X_i$'s, for which the number of records (the black dots) is $R_{24} = 6$. {\bf Right:} Order statistics of $N=7$ random variables. $M_{k,7}$ denotes the $k$-th maximum of the sequence.}\label{fig_record}
\end{figure}

We start by a short review on standard results for record statistics of i.i.d. random variables. We denote by $X_1, X_2, \cdots, X_N$ a collection 
of $N$ i.i.d. random variables, distributed according to a continuous probability density function (pdf) $p(x)$. An entry $X_k$ is an upper record if it is larger than all previous entries (see Fig. \ref{fig_record} left):
\begin{eqnarray}\label{def_record}
X_k > \max\, \{X_1, \cdots, X_{k-1} \} \;, \; k \leq N \;.
\end{eqnarray} 
One can similarly define a lower record which is such that $X_k < {\rm min} \{X_1, \cdots, X_{k-1} \}$. 

In the following, we will focus on upper records (\ref{def_record}), which we will simply call "records". Let $R_N$ be the number of records~(\ref{def_record}) among these $N$ random variables. We first discuss a straightforward method, based on indicator variables, to investigate the statistics of $R_N$. Then we discuss more complicated joint probability distributions of the number and the ages of the records. This second method is not only useful to investigate the age of the largest and smallest record but can be generalized, with some appropriate modifications, to the study of the records of random walks. 
\subsubsection{Distribution of the number of records}

To study this quantity it is useful to introduce indicator variables $\sigma_k$'s which take the value $0$ or $1$:
\begin{eqnarray}\label{def_sigma}
\sigma_k = 
\begin{cases}
& 1 \; {\rm if \;} x_k \; {\rm is \; a \; record} \;, \\
& 0 \; {\rm if \;} x_k \; {\rm is \; NOT \; a \; record} \;
\end{cases}\;, \;
R_N = \sum_{k=1}^N \sigma_k \;.
\end{eqnarray} 
For i.i.d. random variables, these indicator functions $\sigma_k$'s are independent. We define 
\begin{eqnarray}\label{def_rate}
\langle \sigma_k \rangle = r_k \;,
\end{eqnarray}
where the average is taken over the different realizations of the random variables $X_1, \cdots, X_N$: $r_k$ is thus the rate at which a record is broken, at "time" $k$. For i.i.d. random variables, it is straightforward to compute the record rate as it is precisely the probability that the event in Eq. (\ref{def_record}) happens. This yields
\begin{eqnarray}\label{rate_iid}
r_k = \int_{-\infty}^\infty p(x) \left[\int_{-\infty}^x p(y) dy  \right]^{k-1} \, dx = \int_0^1 u^{k-1} du  = \frac{1}{k} \;,
\end{eqnarray}  
where we have used the change of variable $u = \int_{-\infty}^x p(y) dy$. This result $r_k = 1/k$ (\ref{rate_iid}), independently of the parent
distribution, can be easily understood: the probability that $X_k$ is the maximum among $X_1, \cdots, X_k$ is indeed $1/k$ as the maximal value can be realized with equal probability by any of these $k$ i.i.d. random variables. From (\ref{rate_iid}), we get the mean number of records as
\begin{eqnarray}\label{exact_mean_iid}
\langle R_N \rangle = \sum_{k=1}^N r_k = \sum_{k=1}^N \frac{1}{k} = H_{N} \;,
\end{eqnarray}
where $H_{N}$ denotes the $N$-th Harmonic number. For large $N$, it behaves as
\begin{eqnarray}\label{exact_mean_iid_asympt}
\langle R_N \rangle = \log{N} + \gamma_E + {\cal O}(N^{-1}) \;,
\end{eqnarray}
where $\gamma_E = 0.57721\cdots$ is the Euler constant. Similarly, the second moment can be evaluated using indicators variables as
\begin{eqnarray}\label{second_moment}
\langle R_N^2 \rangle - \langle R_N \rangle^2 = \sum_{k=1}^N \langle \sigma_k^2 \rangle - \langle \sigma_k \rangle^2 &=& \sum_{k=1}^N \frac{1}{k} - \frac{1}{k^2} \nonumber \\
&=& \log{N} + \gamma_E - \frac{\pi^2}{6} + {\cal O}(1/N) \;,
\end{eqnarray} 
where we have used, in the first line of Eq. (\ref{second_moment}), that the $\sigma_k$'s are independent. Similarly, one can compute the generating function (GF) of the probability distribution $P(M|N) = {\mathbb P}(R_N = M)$ using (for $N \geq 1$)
\begin{eqnarray}\label{GF_rising}
\sum_{M=1}^\infty P(M|N) x^M &=&  \langle x^{R_N} \rangle = \prod_{k=1}^N \langle x^{\sigma_k} \rangle =  \prod_{k=1}^N \left(\frac{x-1}{k}+1 \right) \nonumber \\
&=& \frac{x(x+1)\cdots(x+N-1)}{N!} \;.
\end{eqnarray}
One recognizes that the rising factorial appearing in (\ref{GF_rising}) is the GF of the unsigned Stirling numbers of the first kind \cite{Riordan}
\begin{eqnarray}
x(x+1)\cdots(x+N-1) = \sum_{M=1}^N {N \brack M} x^M \;,
\end{eqnarray}
where the unsigned Stirling numbers ${N \brack M}$ enumerate the number of permutations of $N$ elements with $M$ disjoint cycles exactly. Hence one has
\begin{eqnarray}\label{exact_dist_number}
P(M|N) = \frac{{N \brack M}}{N!} \;,
\end{eqnarray} 
which thus shows that the number of records of $N$ i.i.d. random variables is distributed like the number of cycles in random permutations of $N$ objects with uniform measure. We will come back later, in section \ref{section:permutation}, to this connection with random permutations. 
Finally, using the asymptotic behaviors of Stirling numbers, one can show that the distribution of $R_N$ approaches, when $N \to \infty$, a Gaussian distribution 
\begin{eqnarray}\label{pdf_iid}
P(M|N) \sim \frac{1}{\sqrt{2 \pi \log N}} \exp{\left(-\frac{(M-\log N)^2}{2 \log N} \right)} \;.
\end{eqnarray}

Here we have discussed the case where the random variables $X_i$'s are continuous random variables. We refer the reader to Ref. \cite{rounding} for a discussion of the effects of discreteness, in particular when continuous random variables are subsequently discretized by rounding to integer multiples of a discretization scale. 

\subsubsection{Joint distribution of the ages of records and of their number}

Let us consider a realization of these $N$ i.i.d. random variables $x_i$'s, which we consider as a time series, the index $i$ playing the role
of discrete time. Let $M$ be the number records in this realization. We denote by ${\vec l} = ({l_1, l_2, \cdots, l_M})$ the time intervals between successive records as depicted in Fig. \ref{fig_record}. Thus $l_i$ is the age of the $i$-th record, i. e. it denotes the time up to which the $i$-th record survives. Note that the last record, the $M$-th record in this sequence, still stays a record at "time" $N$. We first compute the joint probability distribution $P(\vec l, M | N)$ of the ages $\vec l$ and the number $M$ of records, given the length $N$ of the sequence. This joint pdf can be written as
\begin{eqnarray}\label{gen_formula_iid}
&&P(\vec l, M | N) = \int_{-\infty}^\infty dy_M p(y_M) \left[\int_{-\infty}^{y_M} p(x) dx \right]^{l_M-1}  \nonumber \\
&&\times \prod_{k=1}^{M-1} \int_{-\infty}^{y_{k+1}} dy_{k} p(y_k) \left[\int_{-\infty}^{y_{k}} p(x) dx \right]^{l_{k}-1} \, \delta_{\sum_{k=1}^M l_k, N} \;,
\end{eqnarray}
where the delta function in (\ref{gen_formula_iid}) ensures that the size of the sample is $N$. If one performs the change of variables $u_k = \int_{-\infty}^{y_k} p(x) dx$, the pdf $P(\vec l, M | N)$ in (\ref{gen_formula_iid}) can be written as
\begin{eqnarray}\label{gen_formula_iid_2}
P(\vec l, M | N) &=& \int_0^1 du_M u_M^{l_M-1} \prod_{k=1}^{M-1} \int_0^{u_{k+1}} du_{k} u_{k}^{l_{k}-1}\delta_{\sum_{k=1}^M l_k, N}\;.
\end{eqnarray}
This multiple integral in (\ref{gen_formula_iid_2}) can be performed straightforwardly to obtain
\begin{eqnarray}\label{full_jpdf}
P(\vec l, M | N) = \frac{1}{l_1(l_1+l_2)(l_1+l_2+\cdots+l_{M})} \delta_{\sum_{k=1}^M l_k, N}\;. 
\end{eqnarray}
Eq. (\ref{full_jpdf}) carries more information than just the number of records $R_N$. In fact, as we show below, this result in (\ref{full_jpdf}) can be conveniently used to compute the statistics of the age of longest and shortest records.


\subsubsection{Distribution of the age of the longest record}

We now focus on the age of the longest record, denoted by $l_{\max, N}$, which is defined as
\begin{eqnarray}\label{def_lmax}
l_{\max,N} = \max \{l_1, l_2, \cdots, l_M \} \;.
\end{eqnarray}
Its cumulative distribution $F(l |N) = {\mathbb P}(l_{\max,N} \leq l)$, $l \leq 1$, is obtained from the full joint pdf (\ref{full_jpdf}) by summing over $M$ and $l_1, \cdots l_M$ with the constraint that $l_1 \leq l$, $\cdots$, $l_M \leq l$. It reads 
\begin{eqnarray}
F(l | N) = \sum_{M=1}^\infty \sum_{l_1 =1}^l \cdots \sum_{l_M = 1}^l \frac{\delta_{\sum_{k=1}^M l_k, N}}{l_1(l_1 + l_2)\cdots(l_1+l_2+\cdots+l_M)} \;,
\end{eqnarray}
while $F(l | 0) = 1$. The GF of $F(l | N)$ with respect to (wrt) $N$ is conveniently written using the integral representation of the pdf in (\ref{gen_formula_iid_2}) as
\begin{eqnarray}
&&\sum_{N=0} z^N F(l | N) = 1 + \sum_{M=1}^\infty \int_0^1 du_M f(u_M) \prod_{k=1}^{M-1} \int_0^{u_{k+1}} du_k f(u_k) \;, \label{dist_lmax} \\
&& f(u) = z \sum_{m=1}^l (z u)^{m-1} \label{def_f} \;.
\end{eqnarray}
The multiple integral in (\ref{dist_lmax}) can be performed by induction in terms of the integral of $f(u)$
\begin{eqnarray}
g(u) = \int_0^u f(v) dv = \sum_{m=1}^l \frac{(zu)^m}{m} \;, 
\end{eqnarray}
yielding finally
\begin{eqnarray}\label{gf_distlmax}
\sum_{N=0}^\infty z^N F(l | N) = 1 + \sum_{M=1}^\infty \frac{[g(1)]^M}{M !} = \exp{\left(\sum_{k=1}^l \frac{z^k}{k} \right)} \;.
\end{eqnarray}
From the GF of the full distribution of $l_{\max,N}$ (\ref{gf_distlmax}) one obtains the GF of the average value $\langle l_{\max,N} \rangle = \sum_{l=1}^\infty (1 - F(l | N))$ as
\begin{eqnarray}\label{gf_av_lmax}
\sum_{N=0}^\infty \langle l_{\max,N} \rangle z^N = \frac{1}{1-z} \sum_{l=1}^\infty\left[ 1 - \exp{\left(-\sum_{k=l+1}^\infty \frac{z^k}{k} \right)}\right] \;.
\end{eqnarray}
By analysing this expression (\ref{gf_av_lmax}) in the limit $z \to 1$, where the discrete sums can be replaced by integrals (setting $z = e^{-s}$) one obtains
the large $N$ behavior of $\langle l_{\max,N} \rangle$ as
\begin{eqnarray}\label{c1}
\langle l_{\max,N} \rangle = c_1 N + {\cal O}(1) \;, \; c_1 = \int_0^\infty dx (1 - e^{-\int_x^\infty {e^{-y}} \frac{dy}{y}})=0.62432... \;,
\end{eqnarray}
where $c_1$ is the Golomb-Dickman or Goncharov constant \cite{Finch_book}. This constant $c_1$ also describes the linear growth of the longest cycle of a random permutation \cite{Finch_book}. This constant also appeared in a model of growing network \cite{GL2008} and in a one dimensional ballistic aggregation model \cite{MMS09}.

\subsubsection{Distribution of the age of the shortest record}

We now focus on the age of the shortest record, denoted as $l_{\min,N}$, which is defined as
\begin{eqnarray}\label{def_lmin}
l_{\min,N} = \min \{l_1, l_2, \cdots, l_M \} \;.
\end{eqnarray}
We define $G(l | N) = {\mathbb P}(l_{\min,N} \geq l)$, $l \geq 1$, and using the same reasoning as above for $l_{\max,N}$ we find the GF of $G(l | N) = {\mathbb P}(l_{\min,N} \geq l)$ wrt $N$ as
\begin{eqnarray}\label{dist_min_GF}
\sum_{N=0}^\infty G(l|N)z ^N = \exp{\left[\sum_{k=l}^\infty \frac{z^k}{k} \right]} - 1 \;.
\end{eqnarray}
The GF of the average value $\langle l_{\min,N}\rangle = \sum_{l=1}^\infty G(l | N)$ can be obtained from~(\ref{dist_min_GF}) which yields the asymptotic result for large $N$ \cite{SP1966}
\begin{eqnarray}\label{lmin_iid}
\langle l_{\min,N} \rangle = e^{-\gamma_E} \log{N} + o(\log{N}) \;,
\end{eqnarray}
with the numerical value $e^{-\gamma_E} = 0.5614594835...$.

\subsubsection{Connection with random permutations}\label{section:permutation}

As we have seen repeatedly in this section, records statistics bear strong similarities with the statistics of random permutations. The existence of connections between the two fields is well known \cite{Renyi, Goldie} and they recently showed up in various problems of statistical physics \cite{GL2008,MMS09}. One of the main manifestation of this connection is that the number of records $R_N$ for $N$ i.i.d. random variables is distributed like the number of cycles in random permutations of $N$ objects with uniform measure (\ref{exact_dist_number}). We refer the interested reader to Ref. \cite{Flajolet} for a more complete discussion of this connection. 


\subsection{Record statistics of random walks}

We now study the record statistics of a discrete-time random walker (RW) moving on a continuous line. The position of the RW $x_k$ after $k$ steps
evolves via 
\begin{eqnarray}\label{eq:Markov}
x_k = x_{k-1} + \eta_k \;,
\end{eqnarray}
starting from $x_0 = 0$ and where the jump variables $\eta_k$'s are i.i.d. variables, drawn from a distribution $\phi(\eta)$. Here we study the record statistics of a realization of this RW (\ref{eq:Markov}) after $N$ steps, ${x_0, x_1 \cdots, x_N}$ (there are thus $N+1$ random variables in this sequence). As before (\ref{def_record}), a record is broken after $k$ steps if $x_k > \max\{x_0, x_1, \cdots, x_{k-1}\}$, with the convention that $x_0$ is counted as a record. As in the case of i.i.d. variables, we will focus on the number of records $R_N$ as well as on the age of the largest, ${l_{\max,N}}$, and shortest $l_{{\rm min},N}$ record. As discussed before in the case of i.i.d. random variables, the statistics of these quantities are conveniently obtained from the joint probability distribution $P(\vec l, M | N)$ of the ages and the number of records after $N$ time steps. The ages $l_i$'s are thus defined as the number of steps between two records, hence as in the i.i.d. case (see Fig. \ref{fig_record}) except that $l_R \to l_{R}-1$ (in Fig. \ref{fig_record} one would thus have $l_6$ = 5 for a random walk). 

To compute this joint distribution $P(\vec l, M | N)$ we need two quantities as inputs \cite{MZ2008}. The first one is the probability $q_-(l)$ that a RW, starting in $x_0$, stays {\it below} $x_0$ after $l$ time steps:
\begin{eqnarray}\label{def_qminus}
q_-(l) = {\mathbb P}[x_k < x_0 \;, \; \forall \, 1 \leq k \leq l ] \;.
\end{eqnarray}
Due to translational invariance, this probability does not depend on $x_0$ and we can thus set $x_0 = 0$. Its GF, $\tilde q_-(z)$, is given by the generalized Sparre Andersen (SA) theorem \cite{SA54}:
\begin{eqnarray}\label{gen_SA}
\tilde q_- (z) = \sum_{k=0}^\infty q_-(k) z^k = \exp{\left[ \sum_{k=1}^\infty \frac{z^k}{k} {\mathbb P}(x_k < 0)\right]} \;.
\end{eqnarray}
The second quantity is the first passage $f_-(l)$ that the RW crosses its starting point $x_0$ between steps $(l-1)$ and $l$ from below $x_0$. Again, $f_-(l)$ is independent of $x_0$ and one can set $x_0 = 0$. It follows from its definition that $f_-(l) = q_-(l-1) - q_-(l)$ so that its GF can be expressed as
\begin{eqnarray}\label{expr_ftilde}
\tilde f_-(z) = \sum_{l=1}^\infty f_-(l) z^l = 1 - (1-z) \tilde q_-(z) \;.
\end{eqnarray}


Armed with these two quantities $q_-(l)$ and $f_-(l)$ we can then write down explicitly the joint distribution of the ages $\vec l$ and the number of records $M$, $P(\vec l, M | N)$: 
\begin{eqnarray}\label{joint_pdf_rw}
P(\vec l, M | N) = f_-(l_1) f_-(l_2) \cdots f_-(l_{M-1}) q_-(l_M) \delta_{\sum_{k=1}^M l_k,N} \;,
\end{eqnarray}
where we have used the Markov property of the RW which implies that the intervals $l_k$'s are statistically independent, except for an overall global constraint that total length of the interval is $N$, which is incorporated by the delta function. Note that since the number of records is $M$, the last interval $l_M$ is not terminated and its pdf is thus $q_-(l_M)$ instead of $f_-(l_M)$. This exact expression (\ref{joint_pdf_rw}), together with (\ref{gen_SA}) and (\ref{expr_ftilde}) is the starting point of the analysis of record statistics of RW \cite{MZ2008}. It is the analogous to the expression in (\ref{full_jpdf}) obtained in the i.i.d. case.

\subsubsection{Record statistics of a single symmetric random walk}

{\bf Continous jump distribution.} We first consider the case of symmetric jump distributions, such that $\phi(\eta) = \phi(-\eta)$ and focus, for the moment, on the case where  
$\phi(\eta)$ is continuous (the case of lattice RW, with discrete jump distribution, will be discussed below). In this case the SA result (\ref{gen_SA}) becomes, thanks to the fact that ${\mathbb P}(x_k < 0) = 1/2$ for any $k \geq 1$:
\begin{eqnarray}\label{q_symmetric}
\tilde q_-(s)  = \tilde q (s) = \frac{1}{\sqrt{1-s}} \Longrightarrow q_-(l) = q(l) = {2l \choose l} \frac{1}{2^{2l}}  \;,
\end{eqnarray}
independently of the jump distribution $\phi(\eta)$. For large $l$, one has from (\ref{q_symmetric})
\begin{eqnarray}\label{asympt_q_symmetric}
q(l) \underset{l \to \infty}{\sim} \frac{1}{\sqrt{\pi l}} \;.
\end{eqnarray}
On the other hand, from (\ref{joint_pdf_rw}) one gets the GF of $\langle R_N \rangle$ as
\begin{eqnarray}\label{GF_mean_rw}
\sum_{N=0}^\infty \langle R_N\rangle z^N = \frac{1}{(1-z)^2 \tilde q(z)} \;,  
\end{eqnarray}
from which one gets [using (\ref{q_symmetric})]:
\begin{eqnarray}\label{mean_symm_cont}
\langle R_N \rangle = \sum_{k=0}^N {2k \choose k} \frac{1}{2^{2k}} = (2N+1) {2N \choose N} 2^{-2N} \sim \frac{2}{\sqrt{\pi}} \sqrt{N} + {\cal O}(N^{-\frac{1}{2}}).
\end{eqnarray}
It was demonstrated recently that this square root growth $\propto \sqrt{N}$ is robust and remains the same in presence of measurements errors and noise \cite{edery}. Note that from the SA theorem (\ref{q_symmetric}), $q(l)$ and $f_-(l) = f(l) = q(l-1) - q(l)$ are universal, i.e. independent of the jump distribution: hence the full joint distribution $P(\vec l,M|N)$ in (\ref{joint_pdf_rw}) and any of its marginals are also universal. 

Let us first consider the probability distribution of the number of records $P(M|N) = {\mathbb P}(R_N = M) = \sum_{\vec l} P(\vec l,M|N)$. From (\ref{joint_pdf_rw}) one obtains straightforwardly \cite{MZ2008}
\begin{eqnarray}\label{GF_distribution}
\sum_{N=M-1}^\infty P(M|N) z^N = [\tilde f(z) ]^{M-1} \tilde q(z) = \frac{(1-\sqrt{1-z})^{M-1}}{\sqrt{1-z}} \;,
\end{eqnarray}
where we have used (\ref{expr_ftilde}) and (\ref{q_symmetric}). From (\ref{GF_distribution}) it is possible to obtain the full distribution \cite{MZ2008}:
\begin{eqnarray}\label{explicit_dist_continuous}
P(M|N) = {{2N-M+1} \choose {N}} 2^{-2N+M-1} \;, \; M \leq N+1 \;.
\end{eqnarray}
From (\ref{explicit_dist_continuous}) we can obtain the mean as in (\ref{mean_symm_cont}) and the variance, which for large $N$ behaves like
\begin{eqnarray}\label{var_symm_cont}
\langle R_N^2 \rangle - \langle R_N\rangle^2 = 2\left(1 - \frac{2}{\pi} \right) N + {\cal O}(\sqrt{N})\;.
\end{eqnarray}
It is interesting to compare these results for the RW sequence with that of i.i.d. random variables studied above. In particular, for i.i.d. variables, the fluctuations of $R_N$ (\ref{second_moment}) are small compared to the mean (\ref{exact_mean_iid_asympt}) for large $N$. In contrast, for the RW sequence, it follows from (\ref{mean_symm_cont}) and (\ref{var_symm_cont}) that both the mean and the standard deviation grow as $\sqrt{N}$ for $N \gg 1$: thus the fluctuations are large and actually comparable to the mean. This suggests that in the random walk case, at variance with the case of i.i.d. random variables (\ref{pdf_iid}), the probability distribution $F(M|N)$ takes the scaling form $F(M|N) \sim (\sqrt{N})^{-1} g_0(M/\sqrt{N})$. This can actually be shown from the analysis of (\ref{explicit_dist_continuous}) for large $N$, which yields \cite{MZ2008}
\begin{eqnarray}\label{pdf_RN}
P(M |N) \sim \frac{1}{\sqrt{N}} g_0\left(\frac{M}{\sqrt{N}}\right) \;, \; g_0(x) = \frac{1}{\sqrt{\pi}} e^{-\frac{x^2}{4}} \;, \; x > 0 \;.
\end{eqnarray} 

What can be said about the statistics of the ages of the records ? The typical age of record $l_{\rm typ}$ can be estimated as $l_{\rm typ} \sim N/\langle R_N \rangle$, which, from (\ref{mean_symm_cont}), thus grows like $l_{\rm typ} \sim \sqrt{4/\pi} \sqrt{N}$. There are however rare records whose age behaves quite differently with $N$. As was done before in the case of i.i.d. variables we consider the longest lasting record $l_{\max,N}$ in (\ref{def_lmax}) and the shortest duration $l_{{\rm min},N}$ in (\ref{def_lmin}).

We first consider the statistics of $l_{\max,N} = \max \{\, l_1, \cdots, l_M\}$ and compute the cumulative distribution $F(l |N) = {\mathbb P}(l_{\max,N} \leq l)$. As was done before, it can be computed from the full joint pdf $P(\vec l,M|N)$ in (\ref{joint_pdf_rw}) by summing up over $l_i \leq l$ and summing up over $M$. One can thus compute the GF of $F(l |N)$ wrt to $N$ as \cite{MZ2008}
\begin{eqnarray}\label{GF_dist_lmax_rw}
\sum_{N=0}^\infty F(l | N) z^N = \frac{\sum_{k=1}^l q(k) z^k}{1-\sum_{k=1}^l f(k) z^k} \;.
\end{eqnarray}
One can extract, in principle, the expression of $F(l|N)$ from this formula (\ref{GF_dist_lmax_rw}). In particular, the asymptotic large $N$
behavior of the average $\langle l_{\max,N} \rangle = \sum_{l=1}^\infty [1 - F(l|N)]$ can be extracted explicitly \cite{}
\begin{equation}\label{c2}
\langle l_{\max,N} \rangle \sim c_2 N \;, \; c_2 = 2 \int_0^\infty dy \log \left[1 + \frac{1}{2\sqrt{\pi}} \Gamma(-1/2,y)\right] = 0.626508... 
\end{equation} 
Thus the age of the longest record ($\propto N$) is much larger than the typical age ($\propto \sqrt{N}$). Interestingly, the constant $c_2$, for symmetric random walks (\ref{c2}) is quite close to the Golomb-Dickman or Goncharov's constant $c_1$ (\ref{c1}) which characterizes the age of the longest record of a i.i.d. sequence. Note however that the origin of universality is quite different in the two problems. Interestingly, the same constant $c_2$ appears in the excursion theory of Brownian motion \cite{PY97}. The precise link between these two problems was shown in Ref. \cite{GMS2009}. 

For the shortest lasting record $l_{{\rm min},N} = \min \{l_1, \cdots, l_M \}$, it is also useful to consider the cumulative distribution $G(l|N) = {\mathbb P}(l_{{\rm min},N} \leq l)$. Its GF wrt $N$ is easily obtained from the joint pdf (\ref{joint_pdf_rw}) as:
\begin{eqnarray}\label{GF_dist_lmin_rw}
\sum_{N=0}^\infty G(l|N) z^N = \frac{\sum_{k=l}^\infty q(k) z^k}{1- \sum_{k=1}^\infty f(k) z^k} \;.
\end{eqnarray}
In particular, one can extract from (\ref{GF_dist_lmin_rw}) the large $N$ behavior of $\langle l_{{\rm min},N}\rangle$ as \cite{MZ2008}
\begin{eqnarray}
\langle l_{\min, N}\rangle \sim \sqrt{\frac{N}{\pi}} \;,
\end{eqnarray}
which grows in a similar way as that of the typical record, albeit with a smaller prefactor $1/\sqrt{\pi} = 0.56419\cdots$ compared with $\sqrt{\pi/4} = 0.88629 \cdots$. Notice also that it grows much faster ($\propto \sqrt{N}$) than in the case of i.i.d. random variables ($\propto \log N$) (\ref{lmin_iid}).    

\noindent {\bf Discrete lattice random walks.} The above analysis can also be performed for discrete lattice random walks, corresponding to $\phi(\eta) = \frac{1}{2}\delta(\eta+1) + \frac{1}{2}\delta(\eta-1)$ in (\ref{eq:Markov}), except that in this case the expression of $\tilde q(z)$ is different from (\ref{q_symmetric}) for symmetric jump distributions [this can be seen from Eq. (\ref{gen_SA}) as ${\mathbb P}(x_k = 0) \neq 0$ in this case]. 
One has then
\begin{eqnarray}
\tilde q(z) = \tilde q_{-}(z) = \frac{1}{1-z} - \frac{1-\sqrt{1-z^2}}{z(1-z)} \Longrightarrow q(l) \underset{l \to \infty}{\sim}  \frac{\sqrt{2}}{\sqrt{\pi l}} \;,
\end{eqnarray}
which differs, by a factor of $\sqrt{2}$ from (\ref{asympt_q_symmetric}) for the continuous case. In Ref. \cite{MZ2008}, it was shown that $\langle R_N\rangle \sim \sqrt{2N/\pi}$, which is $1/\sqrt{2}$ of the expression for the mean in the continuous case (\ref{mean_symm_cont}). As shown in Ref. \cite{WMS2012}, the number of records $R_N$ is, in this discrete case, directly related to the maximum of the sequence up to step $N$, $M_N = \max{(x_0, \cdots, x_N)}$, via the relation $R_N = M_N + 1$. This allows to compute the full distribution of $R_N$ and show \cite{WMS2012} that for large $N$, it takes the scaling form as in Eq. (\ref{pdf_RN}), with $g_0(x) = \sqrt{2/\pi}e^{-x^2/2}$. Finally, in Ref. \cite{MZ2008}, it was also found that $\langle l_{\max,N}\rangle \sim c_2 N$ and $\langle l_{\min,N}\rangle \sim \sqrt{2N/\pi}$ which are respectively equal to, and $\sqrt{2}$ times, the corresponding expressions for the continuous case.

\subsubsection{Record statistics of a single random walk with a drift}

Up to now, we have discussed the case of symmetric RWs, where the jump length distribution $\phi(\eta)$ is continuous and symmetric, $\phi(\eta) = \phi(-\eta)$. However, the renewal equation for the joint pdf $P(\vec l,M|N)$ in (\ref{joint_pdf_rw}), as well as the generalized SA result (\ref{gen_SA}), are still valid for continuous but asymmetric jump distribution. The only difference is that we have to use the appropriate expressions for $q_-(l)$ (\ref{gen_SA}) and $f_-(l)$ (\ref{expr_ftilde}) instead of $q(l)$ and $f(l)$ in the above expressions for $\langle R_{N} \rangle $ (\ref{GF_mean_rw}) and for the distributions of $l_{\max,N}$ (\ref{GF_dist_lmax_rw}) and $l_{\min, N}$ (\ref{GF_dist_lmin_rw}).

In particular, one can study the case of a biased random walk which is constructed from the symmetric random walk $x_k$ in (\ref{eq:Markov}) as
\begin{eqnarray}\label{RW_drift}
y_k = x_k + c \, k \Longrightarrow y_k = y_{k-1} + c + \eta_k \;,
\end{eqnarray}
where $\eta_k$'s are i.i.d. variables, drawn from a distribution $\phi(\eta)$: $y_k$ thus represents the position of a discrete-time random walker at step $k$ in presence of a constant drift $c$. In Ref. \cite{LDW09}, the authors studied in detail the special case of the Cauchy jump density, $\phi_{\rm Cauchy}(\eta) = 1/[\pi(1+\eta^2)]$ with arbitrary drift $c$. In particular it was found that the mean number of records depends algebraically on $N$ with a continuously varying exponent $\theta(c)$ \cite{LDW09}
\begin{eqnarray}\label{theta_c}
\langle R_N \rangle \sim N^{\theta(c)} \;, \; \theta(c)= \frac{1}{2} + \frac{1}{\pi} {\rm arctan}{(c)} \;.
\end{eqnarray} 
On the other hand, the mean number of records $\langle R_N \rangle$ for jump densities with a finite second moment $\sigma^2$ and positive drift $c > 0$ was studied in Ref.~\cite{WBK2011}, using various approximation schemes. In Ref. \cite{MSW2012} the authors studied the record statistics of such a biased random walk (\ref{RW_drift}) for arbitrary continuous jump distribution $\phi(\eta)$ such that its Fourier transform $\hat \phi(q) = \int_{-\infty}^\infty \phi(\eta) e^{i q \eta} d\eta$ behaves, for small $q$, as 
\begin{eqnarray}\label{def_mu}
\hat \phi(q) = 1 - |l_\mu q|^\mu + o(|q|^\mu) \;,
\end{eqnarray}
where $0 < \mu \leq 2$ and $l_\mu$ is a typical length scale of the jump. The exponent $\mu$ controls the large $|\eta|$ tail of $\phi(\eta)$. For jump density with a well defined second moment $\sigma^2 = \int_{-\infty}^\infty d\eta \eta^2 \phi(\eta)$ one has evidently $\mu = 2$, while for $\mu \in (0,2)$, $\phi(\eta)\sim |\eta|^{-1-\mu}$ for large $|\eta|$. The record statistics of such RW (\ref{RW_drift}) for any value of $\mu \in (0,2]$ (\ref{def_mu}) and any drift $c$ was performed in Ref. \cite{MSW2012}. The analysis performed relied on a detailed study of the behavior of the persistence probability $q_-(l)$ which was found to be very sensitive to these parameters $\mu$ and $c$. This study \cite{MSW2012} revealed the existence of five distinct regions in the $(c, 0 < \mu \leq 2)$ strip where $R_N$, $\langle l_{\max,N} \rangle$ and $\langle l_{\max,N} \rangle$ exhibit very different behaviors. These results are summarized in Table 1.

\begin{table}[hh]\label{table}
\tbl{Summary of the main results for the records of RWs with a drift (\ref{RW_drift}), from Ref. \cite{MSW2012}. The constant $c_2$ is given in (\ref{c2}) and the exponent $\theta(c)$ is given (\ref{theta_c}), while the constants $0< a_\mu(c) < 1$ and $0< C_{\rm II} < 1$ are non-universal constants given in \cite{MSW2012}.}{
\begin{tabular}{|c||c|c|c|c|c|}
\hline
\quad & \quad & \quad & \quad & \quad & \quad \\
 \quad & {\bf I} & {\bf II} & {\bf III} & {\bf IV} & {\bf V} \\
 \quad & $\mu \in (0,1)$  & $\mu = 1$ & $\mu \in (1,2)$ & $\mu = 2$ & $\mu \in (1,2]$ \\
 \quad & $c \in \mathbb R$  & $c \in \mathbb R$ & $c \in \mathbb R^+$ & $c \in \mathbb R^+$ & $c \in \mathbb R^-$ \\
 \quad & \quad & \quad & \quad & \quad & \quad \\
\hline
\hline
\quad & \quad & \quad & \quad & \quad & \quad \\
$q_-(N)$ & $\propto  N^{-\frac{1}{2}}$ & $\propto  N^{-\theta(c)}$ & $\propto  N^{-\mu}$ & $\propto N^{-\frac{3}{2}} e^{-\frac{c^2 N}{2\sigma^2}}$ & $\sim a_\mu(-c)$ \\
\quad & \quad & \quad & \quad & \quad & \quad \tabularnewline
\hline
\quad & \quad & \quad & \quad & \quad & \quad \\
$\langle R_N \rangle $ & $\propto  {N}^{\frac{1}{2}}$ & $\propto  N^{\theta(c)}$ & $\sim a_\mu(c) N$ &  $\sim a_2(c) N$ & $\sim [a_{\mu}(-c)]^{-1}$ \\
\quad & \quad & \quad & \quad & \quad & \quad \\
\hline
\quad & \quad & \quad & \quad & \quad & \quad \\
$\langle l_{\max, N} \rangle$ & $\sim c_2 \, N$ & $\sim C_{\rm II} \, N$ & $\propto  N^{\frac{1}{\mu}}$ & $\propto \log N$ & $\sim N$ \\
\quad & \quad & \quad & \quad & \quad & \quad \\
\hline
\quad & \quad & \quad & \quad  & \quad & \quad \\
$\langle l_{\min, N} \rangle$ & $\propto  {N}^{\frac{1}{2}}$ & $\propto N^{1-\theta(c)}$ & $\sim 1-a_{\mu}(c)$ & $\sim 1-a_2(c)$ & $\sim a_\mu(-c) N$ \\
\quad & \quad & \quad & \quad  & \quad & \quad \\
\hline
\end{tabular}}
\caption{Summary of our results for the three different models A, B and C.}\label{table}
\end{table}

\subsubsection{Record Statistics for Multiple Random Walks}

We conclude this section on records by mentioning results for the records of $n$ symmetric independent RW's, which were obtained in \cite{WMS2012}. 
At each time step, each walker jumps by a random length drawn independently from a 
symmetric and continuous distribution, as in (\ref{eq:Markov}). Two cases were considered: (I) when the variance $\sigma^2$
of the jump distribution is finite and (II) when $\sigma^2$ is divergent as in
the case of L\'evy flights with index $0 < \mu
< 2$~(\ref{def_mu}).  
In both cases it was found that the mean record number $\langle R_{N,n} \rangle$ grows 
universally as $\sim \alpha_n \sqrt{N}$ for large $N$, but with a very different behavior 
of the amplitude $\alpha_n$ for $n > 1$ in the two cases. Indeed it was shown that, for large $n$, $\alpha_n \approx 2 
\sqrt{\log n} $ independently of $\sigma^2$ in case I while, in case II,
the amplitude approaches to an $n$-independent constant for large $n$,
$\alpha_n \approx 4/\sqrt{\pi}$, independently of $0<\mu<2$.
For finite 
$\sigma^2$ it was argued, and this was confirmed by numerical simulations, that the 
full distribution of $(R_{N,n}/\sqrt{N} - 2 \sqrt{\log n}) \sqrt{\log n}$ converges to a 
Gumbel law [as in Eq. (\ref{Gumbel}) below] as $N \to \infty$ and $n \to \infty$. In case II, numerical simulations 
indicated that the distribution of $R_{N,n}/\sqrt{N}$ converges, for $N \to \infty$ and $n 
\to \infty$, to a universal nontrivial distribution, independently of $\mu$, the computation of which remains an open 
problem. Ref. \cite{WMS2012} also discussed 
the applications of these results on records for multiple random walks to the study of the record statistics of 366 daily stock 
prices from the Standard \& Poors 500 index.

\section{Order statistics}

\subsection{Order statistics of i.i.d. random variables}

Let us first review the standard results for order statistics of i.i.d. random variables. We refer the reader to classical textbooks \cite{ABN1992,DN2003} on the subject for more details (see also Ref. \cite{bertin} for a review). We denote by $X_1, X_2, \cdots, X_N$ a collection 
of $N$ i.i.d. random variables, distributed according to a probability density function (pdf) $p(x)$. We denote their common cumulative distribution by $P(x) = \int_{-\infty}^x p(y) dy$. We define the $N$ order statistics of this sequence by arranging the values of $X_i$ by decreasing order of magnitude (see Fig. \ref{fig_record} right)
\begin{eqnarray}\label{def_order}
X_{\max} = M_{1,N} > M_{2,N} > \cdots > M_{N,N} = X_{\min} \;,
\end{eqnarray}  
where we denote by $X_{\max}$ and $X_{\min}$ the maximum and the minimum among the $X_i$'s. 

For i.i.d. random variables, it is possible to write down explicitly the full joint distribution $p_N(m_1, \cdots, m_N)$ of $M_{1,N}, \cdots, M_{N,N}$. To compute it, we first note that, given the realizations of the $N$ order statistics to be $m_1 > m_2 \cdots > m_N$, the original variables $X_i$'s are constrained to take on the values $m_i$ ($i=1,2, \cdots, N$). On the other hand, by symmetry, each of the $N !$ permutations of the $X_i$'s are assigned
the same weight. Hence we have~\cite{ABN1992,DN2003}
\begin{eqnarray}\label{joint_order_iid}
p_N(m_1, \cdots, m_N) = N ! \prod_{i=1}^N p(m_i) \prod^{N-1}_{i=1}{\theta(m_i - m_{i+1})}\;,
\end{eqnarray} 
where the product of $\theta$ functions ensures the ordering of the variables (we remind that $\theta(x) = 1$ if $x>0$ while $\theta(x) = 0$ if $x<0$). From this expression (\ref{joint_order_iid}) one can get, in principle, any characteristic of order statistics of i.i.d. random variables. Here, in addition to the distribution of $M_{k,N}$ we study the gap between two successive maxima (see Fig. \ref{fig_record} right)
\begin{eqnarray}\label{def_gap}
d_{k,N} = M_{k,N} - M_{k+1,N} \;,
\end{eqnarray}
which is an interesting characteristic of the crowding near extreme events \cite{SM07,SMR2008}. 
 
\subsubsection{Finite sample}

We first focus on the pdf $f_{k,N}(m) = \partial_m {\mathbb P}(M_{k,N} \leq m)$ which can be obtained from the full joint pdf (\ref{joint_order_iid}) by integrating over $m_1 \cdots, m_{k-1}, m_{k+1}, \cdots, m_N$ such that $m_1>\cdots>m_{k-1}>m>m_{k+1} > \cdots > m_{N}$:
\begin{eqnarray}\label{pdf_k_iid_step1}
f_{k,N}(m) &=& N ! \, p(m) \int_m^\infty d m_{k-1} p(m_{k-1}) \prod_{j=1}^{k-2} \int_{m_{j+1}}^\infty dm_j p(m_j) \nonumber \\
&\times& \int_{-\infty}^m dm_{k+1} \prod_{j=k+2}^N \int_{-\infty}^{m_{j-1}} dm_j p(m_j) \;,
\end{eqnarray}
where $P(m) = \int_{-\infty}^m p(m') dm'$. It is then straightforward to check (for instance by induction) that $f_{k,N}$ in~(\ref{pdf_k_iid_step1}) can be written as:
\begin{eqnarray}\label{exact_pdf_k_iid}
f_{k,N}(m) = \frac{N!}{(k-1)! (N-k)!} p(m) \left[P(m) \right]^{N-k}  \left[1 - P(m)\right]^{k-1} \;.
\end{eqnarray}
Note that this formula (\ref{exact_pdf_k_iid}) can also be directly obtained by noticing that the event that $m < M_{k,n} < m + dm$ is the same as the following event where $X_i \geq m+dm$ for $(k-1)$ of the $X_i$'s, $m < X_i < m + dm$ for exactly one of the $X_i$'s and $X_i \leq m$ for the remaining $N-k$ $X_i$'s. In particular for $k=1$ one obtains from (\ref{exact_pdf_k_iid}) the pdf of $X_{\max}$ as
\begin{eqnarray}
f_{1,N} = \frac{\partial}{\partial m} {\mathbb P}(X_{\max} \leq m) = N p(m) \left[P(m)\right]^{N-1} \;.
\end{eqnarray} 
Similarly the pdf of $X_{\min}$ is obtained by setting $k=N$ in (\ref{exact_pdf_k_iid}). 

From the full joint pdf in (\ref{joint_order_iid}) it is also possible to obtain the joint pdf of $M_{j,N}$ and $M_{k,N}$ and eventually the pdf $p_{k,N}(d)$ of the gap $d_{k,N} = M_{k,N} - M_{k+1,N}$ as
\begin{eqnarray}\label{dist_gap_iid}
p_{k,N}(d) = \theta(d) \frac{N!}{(N-k-1)!(k-1)!} && \int_{-\infty}^\infty dm_k p(m_k) p(m_{k}-d)  \\
&\times& \left[P(m_{k}-d)\right]^{N-k-1} \left[ 1 - P(m_k)\right]^{k-1} \;. \nonumber
\end{eqnarray}
As an example, for the case of exponential i.i.d. random variables $X_i$'s, such that $p(x) = \theta(x) e^{-x}$, one finds from (\ref{dist_gap_iid}) 
\begin{eqnarray}\label{gap_exp_iid}
p_{k,N}(d) = \theta(d) k e^{-k d} \;,
\end{eqnarray}
independently of $N$.

\subsubsection{Asymptotic results for large samples}

We now turn to the analysis of these results for i.i.d. variables in the limit of large samples, where $N$ is large. We first focus on $F_{1,N}(m) = {\mathbb P}(X_{\max} \leq m)$ which is known to exhibit a universal behavior when $N \to \infty$. Indeed, one can show that there exist constants $a_N$ and $b_N$ and three distinct families of distributions $G_\rho(z)$, $\rho={\rm I},{\rm II},{\rm III}$ such that
\begin{eqnarray}
\lim_{N \to \infty} F_{1,N}(a_N + b_N z) \to G_{\rho}(z) \;, \; \rho = {\rm I},{\rm II}, {\rm III} \:,
\end{eqnarray}
where the limiting distribution $G_\rho(z)$, depends only the large argument of the parent distribution $p(x)$. The large $N$ behavior of extreme value statistics (EVS) of i.i.d. variables is thus characterized by three distinct universality classes: (I) Gumbel, (II) Fr\'echet and (III) Weibull.

{\bf The Gumbel universality class}. In this case, the support of $p(x)$ might be bounded or unbounded -- though the later is the most commonly encountered. In that case, the Gumbel universality class corresponds to the case where $p(x)$ decays faster than any power law, $p(x) \ll x^{-\eta}$, for any value of $\eta>0$ and $G_{\rm I}(z)$ is given by a double exponential
\begin{eqnarray}\label{Gumbel}
G_{\rm I}(z) = \exp{\left[- \exp{(-z)}\right]} \;,
\end{eqnarray}
the so called Gumbel distribution. The constant $a_N$ is given by the standard relation of EVS
\begin{eqnarray}
1 - P(a_N) = \int_{a_N}^\infty p(x) dx = \frac{1}{N} \;,
\end{eqnarray}
which simply says that there is typically one single variable, the maximum, in the interval $[a_N,+\infty)$. On the other hand 
$b_N$ is given by the relation 
\begin{eqnarray}
b_N = \frac{\int_{a_N}^\infty (x-a_N)p(x) dx}{\int_{a_N}^\infty p(x) \, dx} \;,
\end{eqnarray}  
which can be interpreted as the typical distance between $X_{\max}$ and $a_N$, conditioned to the fact that there is a single variable in $[a_N,+\infty)$. The Gumbel universality class corresponds to the case where, for instance, $p(x)$ is an exponential or a Gaussian distribution. But this also corresponds to the case where $p(x)$ is defined on a bounded support, for instance $x \in [0,1)$ where $p(x)$ exhibits an essential singularity in $x=1$, $p(x) \sim \exp{\left[-1/(1-x)^\nu \right]}$, with $\nu > 0$.   

{\bf The Fr\'echet universality class.} This class corresponds to the case where the support of $p(x)$ is unbounded and where $p(x)$ has a power law
tail $p(x) \propto x^{-1-\alpha}$, with $\alpha > 0$. In this case the limiting distribution $G_{\rm II}(z)$ is given by
\begin{eqnarray}\label{Frechet}
G_{\rm II}(z) = \theta(z) \exp{\left[ - z^{-\alpha}\right]} \;.
\end{eqnarray}
Besides, one has in this case $a_N = 0$ while $b_N$ is given by
\begin{eqnarray}\label{bn_frechet}
1 - P(b_N) = \int_{b_N}^\infty p(x) dx = \frac{1}{N} \;,
\end{eqnarray}
from which one gets in particular that $b_N \propto N^{\frac{1}{\alpha}}$. This situation corresponds to the case where $p(x)$ is, for instance, a Cauchy distribution or a Pareto distribution. 

{\bf The Weibull universality class.} This corresponds to the situation where the support of $p(x)$ is bounded from above, such that $p(x) = 0$ if $x>x^*$ and $p(x)$ behaves when $x$ approaches $x^{*}$ as $p(x) \propto (x^* - x)^{\alpha-1}$, $\alpha > 0$. In this case the limiting distribution $G_{\rm III}(z)$ is given by
\begin{eqnarray}\label{Weibull}
G_{\rm III}(z) =
\begin{cases}
&1 \;, \; z > 0 \;, \\
&\exp{\left[-|z|^\alpha \right]} \;, \; z < 0 \;.
\end{cases}
\end{eqnarray}   
In this third case, one has naturally $a_N = x^*$ while $b_N$ is given by 
\begin{eqnarray}
\int_{x^*-b_N}^{x*} p(x) dx = \frac{1}{N} \;,
\end{eqnarray}
from which one gets that $b_N\propto N^{-\frac{1}{\alpha}}$. This universality class includes, for instance, the case where $p(x)$ is a uniform distribution, $p(x) = \theta(x) \theta(1-x)$ (and in this case $\alpha = 1$). 

One can now study the limiting behavior of the distribution of the $k$-th maximum $F_{k,n}(m)$. In this case, depending on the parent distribution $p(x)$, which might belong to one of the three aforementioned universality classes, $\rho= 1,2,3$, one can show that \cite{ABN1992,DN2003}
\begin{eqnarray}
\lim_{N \to \infty} F_{k,N}(a_N + b_N z) &=& G_{\rho}(z) \sum_{j=0}^{k-1} \frac{\left[-\log G_{\rho}(z)\right]^j}{j!} \;, \\
& =& \frac{1}{(k-1)!} \int_{[-\ln G_{\rho}(z)]}^\infty e^{-t} t^{k-1} dt \;,
\end{eqnarray}
where $G_\rho$, with $\rho = 1, 2, 3$, is one of the three limiting distributions mentioned above in Eqs. (\ref{Gumbel}), (\ref{Frechet}) or (\ref{Weibull}). 

A more complete result can also be obtained for the full asymptotic distribution of the vector of the $k$ first maxima \cite{ABN1992,DN2003}
\begin{eqnarray}
\left(\frac{M_{1,N} - a_N}{b_N},  \frac{M_{2,N} - a_N}{b_N}, \cdots, \frac{M_{k,N} - a_N}{b_N}\right) \underset{N \to \infty}{\longrightarrow} (W_1, \cdots, W_k) 
\end{eqnarray}
where the joint pdf of $W_1, \cdots, W_k$ is given by
\begin{eqnarray}\label{limiting_joint_iid}
p(w_1, \cdots, w_k)= G_{\rho}(w_k) \prod_{i=1}^k \frac{g_\rho(w_i)}{G_\rho(w_i)} \;,\; w_1 >\cdots > w_k \;,
\end{eqnarray}
where $g_\rho(z) = G_\rho'(z)$. This expression (\ref{limiting_joint_iid}) is already well known. From it we derive the expression for the  
limiting distribution of the $k$-th gap $d_{k,N} = M_{k,N}- M_{k+1,N}$, which we have not seen in the literature before. It reads:
\begin{eqnarray}
p_{k,N}(d) \sim \frac{1}{b_N} p_{{\rm gap},\rho}\left(\frac{d}{b_N}\right) \;,
\end{eqnarray}
where $p_{\rm gap,\rho}(d)$ is given by
\begin{eqnarray}\label{asympt_gap_iid}
p_{\rm gap,\rho}(d) = \frac{\theta(d)}{(k-1)!} \int_{-\infty}^\infty g_\rho(x) \frac{g_{\rho}(d+x)}{G_\rho(d+x)} \left[ - \log G_\rho(d+x)\right]^{k-1} \; dx \;.
\end{eqnarray}
In particular, for the Gumbel universality class, one finds simply
\begin{eqnarray}
p_{\rm gap, \rm I}(d) = \theta(d) k e^{- k d} \;. 
\end{eqnarray}
For the Fr\'echet universality class, i.e. $\rho = 2$, one finds from (\ref{asympt_gap_iid}):
\begin{eqnarray}\label{gap_frechet}
p_{\rm gap, \rm II}(d) = \theta(d)\frac{\alpha^2}{(k-1)!} \int_0^\infty e^{-x^{-\alpha}} x^{-\alpha-1} (x+d)^{-\alpha k -1} dx \;.
\end{eqnarray} 
In particular for large $d$, it behaves like
\begin{eqnarray}\label{gap_frechet_large}
p_{\rm gap, \rm II}(d) \underset{d \to \infty}{\sim} \frac{\alpha}{(k-1)!} d^{-\alpha k - 1 } \;.
\end{eqnarray}
For $\alpha = 1$, the above integral (\ref{gap_frechet}) can be explicitly evaluated
\begin{eqnarray}
p_{\rm gap, \rm II}(d) = \theta(d) k(k+1) d^{-1-k} {\rm U}(k+1,0,1/d) \;,
\end{eqnarray}
where ${\rm U}(a,b,z)$ is the confluent (Tricomi) hypergeometric function, which is consistent, for large $d$ with (\ref{gap_frechet_large}) for $\alpha = 1$. 

Finally, for the Weibull universality class, one finds
\begin{eqnarray}\label{gap_weibull}
p_{\rm gap, \rm III}(d) = \theta(d)\frac{\alpha^2}{(k-1)!} \int_0^\infty dx (x+d)^{\alpha-1} e^{-(x+d)^\alpha} x^{\alpha k -1} \;,
\end{eqnarray}
which for large $d$ behaves like
\begin{eqnarray}
p_{\rm gap, \rm III}(d) \underset{d \to \infty}{\sim} \frac{\alpha^2}{(k-1)!} \alpha^{- k \alpha} \Gamma(\alpha k) d^{(1-\alpha)(\alpha k - 2)} e^{-d^\alpha} \;.
\end{eqnarray}
For $\alpha=1$, this expression (\ref{gap_weibull}) simplifies to yield simply
\begin{eqnarray}
p_{\rm gap, \rm III}(d) = \theta(d) e^{-d} \;.
\end{eqnarray}

\subsection{Order statistics of random walks}

As we have seen, the order statistics of i.i.d. random variables is fully understood, thanks in particular to the identification
of three different universality classes. In this section, we present recent results for the order statistics of random walks, which 
offer a non-trivial instance of a set of strongly correlated variables where exact results can be obtained. We will see that the results
are quite different from the i.i.d. case. 

We thus consider a RW which starts at $x_0=0$ at time $0$ and evolves via (\ref{eq:Markov}) 
where the $\eta_k$'s are i.i.d. random jumps each drawn from a symmetric distribution $\phi(\eta)$. We study the fluctuations of the ordered sequence $M_{1,N} > M_{2,N} > \cdots >  M_{N+1,N}$ where $M_{k,N}$ is the $k$-th maximum of the RW after $N$ time steps, hence $k=1, \cdots, N+1$.
The study of order statistics for random walks, beyond the first maximum $X_{\max} = M_{1,N}$, was initiated recently in Ref. \cite{SM12} for the case where the jump distribution $\phi(\eta)$ has a well defined second moment $\sigma^2$. In this case, the RW converges, in the limit of a large number of steps $N$, to the  Brownian motion. In Ref. \cite{SM12}, it was shown in this case that when $N \to \infty$
\begin{eqnarray}
\frac{\langle M_{k,N}\rangle}{\sigma} = \sqrt{\frac{2N}{\pi}} + {\cal O}(1) \;,
\end{eqnarray}
independently of $k$. Thus the property of the crowding of extremum ($k$-dependence) 
is not captured by the statistics of the maxima $M_{k,N}$ themselves,
at least to leading order for large $N$. The simplest observable
that is sensitive to the crowding phenomenon is thus the gap,
$d_{k,N} = M_{k,N} - M_{k+1,N}$. The main result of Ref. \cite{SM12} is to show that the
statistics of the scaled gap $d_{k,N}/\sigma$ becomes stationary, 
i.e., independent of $N$ for 
large $N$, but
retains a rich, nontrivial $k$ dependence which becomes {\em 
universal} for large $k$, i.e. independent of the details of the jump 
distribution $\phi(\eta)$.

In particular, using the so called Pollaczek-Wendel identity \cite{Pol75,wendel}, the stationary mean gap $\bar d_k=\langle d_{k,\infty}\rangle$  was
computed exactly for all $k$ and for arbitrary $\phi(\eta)$ [whose Fourier transform is denoted by $\hat \phi(q)$] \cite{SM12}
\begin{eqnarray}
\bar d_k=\langle d_{k,\infty}\rangle =  \frac{\sigma}{\sqrt{2 \pi}} \frac{\Gamma(k+\frac{1}{2})}{\Gamma(k+1)}   - \frac{1}{\pi k} \int_0^\infty \frac{dq}{q^2} \left[ [\hat \phi (q) ]^k - \frac{1}{(1 + \frac{\sigma^2}{2}q^2)^k} \right]. \label{exact_gap}
\end{eqnarray}
In the limit of large $k$, one finds from (\ref{exact_gap}) that
\begin{eqnarray}\label{dbark}
\frac{\bar d_k}{\sigma} \sim \frac{1}{\sqrt{2 \pi k}} \;,
\end{eqnarray}
independently of $\phi(\eta)$. This $k^{-1/2}$ dependence in $\bar d_k$ (\ref{dbark}) was actually noticed in the numerical study of
periodic random walks in Ref. \cite{MOR2011} and was also conjectured to be exact, based on scaling arguments. 

It is natural to wonder about the full distribution of the stationary gap, not only its first moment (\ref{exact_gap}). In Ref. \cite{SM12}, this full pdf $p_k(\delta) d\delta={\mathbb P}(d_{k,\infty} \in [\delta, \delta + d\delta])$ was computed exactly, using backward Fokker-Planck techniques \cite{CM05}, for one particular case of a jump variables with an exponential distribution $\phi(\eta)=b^{-1}\exp\left(-|\eta|/b\right)$. In the limit of large $k$,
it was shown that there is a scaling regime when 
$\delta\sim \langle
d_{k,\infty}\rangle\simeq \sigma/
\sqrt{2\pi k}$ where the pdf scales as, $p_k(\delta)\simeq 
(\sqrt{k}/\sigma) P(\delta
\sqrt{k}/\sigma)$, with a nontrivial scaling function
\begin{equation}\label{exact_F} 
P(x) = 4\big[\sqrt{\frac{2}{\pi}}(1+2x^2) -
e^{2x^2}x(4x^2+3) {\rm erfc}(\sqrt{2}x)\big] 
\,, 
\end{equation} 
where ${\rm
erfc}(z) = (2/\sqrt{\pi})\int_z^\infty e^{-t^2} \, dt$ is the complementary   
error function. While it was not possible to compute the gap pdf for
arbitrary $\phi(\eta)$, numerical simulations \cite{SM12} provided
strong evidence that the scaling function $P(x)$ in Eq. (\ref{exact_F})
is actually universal, i.e., independent of $\phi(\eta)$. 
Somewhat unexpectedly, we find that this universal scaling function 
has an algebraic tail $P(x)\sim x^{-4}$ for large $x$.
For $\delta \gg \langle
d_{k,\infty}\rangle\simeq \sigma/
\sqrt{2\pi k}$, the pdf gets cut-off in a nonuniversal fashion. Thus there are two scales associated to  $d_{k,\infty}$: a typical fluctuation
which is universal and large fluctuations which are non-universal. This is shown to have interesting consequences for the
moments of the stationary gap:
$\langle d_{k,N}^p \rangle \sim 
k^{-\frac{p}{2}}$ for $p<3$,
while $\langle d_{k,N}^p \rangle \sim k^{-\frac{3}{2}}$ for $p>3$.

We end up this section on order statistics of RW by mentioning that exact results have been recently obtained, using first-passage techniques,  
for the joint distribution $P_N(g,l)$ of the first gap $d_{1,N} = G_N = M_{1,N} - M_{2,N}$ and the time $L_N = n_1 - n_2$ between the occurrence of these first two maxima \cite{MMS13}. This analysis was carried out for any value of the L\'evy index $0 < \mu \leq 2$ (\ref{def_mu}).
In particular, it was shown that $P_N(g,l)$ converges to a stationary distribution, i.e. independent of $N$ for large $N$, which displays a very rich behavior as a function of $g$ and $l$ as $\mu$ is varied.

\section{Conclusion}

To conclude, after a brief review on records and order statistics for i.i.d. random variables, we have presented the main results which were recently obtained for records and order statistics of RW, using first-passage concepts. A striking feature of these statistics for $N$ i.i.d. random variables is their universality with respect to their common parent distribution. For records, universality shows up, to a large extent, already for any finite $N$. This can be seen, for instance, through their connection with the statistics of random permutations. For extreme and order statistics, universality only appears in the (thermodynamical) limit of large $N$, thanks to the existence of three distinct universality classes (Gumbel, Fr\'echet and Weibull). What is left of this universal behavior in the presence of strong correlations is an important question. Quite interestingly, for RW with symmetric and continuous jump distribution $\phi(\eta)$, the records statistics do not depend on the details of $\phi(\eta)$ (including L\'evy RW such that $\phi(\eta) \sim |\eta|^{-1-\mu}$ with $0<\mu < 2$), even for a finite number of steps. This universality is due here to the Sparre Andersen theorem. In the presence of a drift $c$, universal behavior also emerges but only in the limit of a large number of steps $N \to \infty$. However in this case, this asymptotic behavior depends on both $c$ and the L\'evy index $\mu$ (see Table 1). 

On the other hand, order statistics of RW is quite sensitive to the jump distribution $\phi(\eta)$. For instance, the distribution of the gap $d_{k,N}$ for finite $k$ and $N$ is generically quite sensitive on $\phi(\eta)$. However, for large $k$ and large $N$, a scaling regime was identified when $d_{k,N}\sim1/\sqrt{k}$ where the fluctuations are universal and described by a universal scaling function (\ref{exact_F}), at least in the case where the jump distribution $\phi(\eta)$ has a finite second moment. The statement of the universality of this scaling regime is based on (i) exact calculation for the case of exponential jumps, (ii) numerical simulations. It will be interesting to establish this universal behavior on firmer grounds. Finally, it will be interesting to extend this study of records and order statistics to other stochastic processes, in particular non-Markovian ones.

\end{document}